\documentclass{article}

\usepackage{graphicx}
\usepackage{float}
\usepackage[a4paper, top=1.5cm, bottom=2.5cm, left=3cm, right=3cm]{geometry}

\usepackage[
  backend=biber,
  style=numeric-comp,
  sorting=none
]{biblatex}

\usepackage{authblk}
\usepackage{marvosym} 

\date{} 

\usepackage{xcolor}
\usepackage{fancyhdr}
\usepackage{amsmath}
\usepackage{lastpage}

\newcommand{\arxiv}{ar\textcolor{black}{X}\raisebox{0ex}iv}

\fancypagestyle{firstpage}{
  \fancyhf{}

  \fancyfoot[C]{%
    \small
    Witte\ \textbar\ 
    \arxiv\ \textbar\ 
    January 2025\ \textbar\ 
    1--\pageref{LastPage}
  }
}

\title{Computation as Organisation}

\author{Kimia Witte}

\affil{Department of Biomedical Engineering, University of Strathclyde, Glasgow}

\begin{document}
\maketitle
\thispagestyle{firstpage}

\section{Abstract}

Computation is commonly defined as the execution of abstract algorithms over symbolic representations, with physical systems treated as substrates that realise predefined operations. While effective for engineered machines, this separation becomes problematic when applied to living systems, where persistence, adaptation, and failure occur without symbolic instruction or central control. Here, computation is reformulated as a structural property of organised matter. Organisation is defined as the persistence of relational constraints that delimit admissible state transitions. Information is not encoded content but relational invariance: differences that influence future behaviour by reshaping what transitions remain possible. Computation is identified with the ongoing enactment of such organisation, integrating memory, processing, and execution as inseparable aspects of material dynamics. Within this framework, algorithms correspond to internally embedded regularities enabled by constraint, and computational limits arise from organisation itself. The account provides experimentally accessible criteria for computation based on persistence, recovery, and structural failure under perturbation.

\vspace{0.25 in}

\noindent \textbf{\small {Keywords: Computation as organisation; relational invariance; constraint-based dynamics; non-representational computation; organisational persistence}}

\section{Introduction}

Contemporary theories of computation typically distinguish abstract algorithms from the physical systems that realise them. Within this framework, computation is specified at a logical level, while physical systems are treated as substrates that instantiate predefined operations. Algorithms are defined independently of material context, and physical systems are said to compute insofar as their dynamics correspond to an abstract specification defined elsewhere. This separation between logical description and physical realisation underpins much of contemporary computer science.

This separation permits generalisation across machines and supports formal analysis. It also entails an ontological assumption: that computation is distinct from physical interaction itself. Matter, on this view, does not compute by virtue of its dynamics, but only insofar as those dynamics correspond to an abstract specification defined elsewhere.

When this assumption is applied to biological systems, conceptual difficulties arise. Living systems exhibit persistence, robustness, adaptation, and failure without executing symbolic programs or maintaining explicit representations. Cellular responses to perturbation do not rely on encoded instructions; tissue-level coordination does not require central control; organismal coherence is maintained without access to a complete system state. In such systems, memory, processing, and execution are not separable stages but occur concurrently through material interaction.

Describing these systems as implementing computation introduces representational categories whose physical referents are often unclear. Terms such as 
"information processing" or "decision-making" are frequently applied without specifying what constitutes a symbol, an instruction, or an execution step in material terms. As a result, computation becomes a descriptive label applied post hoc rather than a property grounded in identifiable physical criteria.

Several approaches have attempted to address this mismatch by expanding the scope of computation. Frameworks such as natural computation, morphological computation, and biology-first approaches emphasise embodiment, interaction, and constraint. These approaches reduce reliance on symbolic representation, but typically retain a conceptual distinction between computation and the material processes that realise it. Computation remains something attributed to matter rather than identified with it.

This identification has several consequences. It removes the need for representational intermediaries between matter and computation. No computational privilege is associated with life: systems compute insofar as their organisation persists under constraint, and non-living systems may compute transiently or locally under the same criteria. The distinction is not between kinds of systems, but between persistence and collapse of organisation. As a result, limits of computation are situated within organisational structure rather than formal abstraction alone. The identification also provides criteria by which computation can be investigated experimentally, not through input–output correspondence, but through persistence, recovery, and structural failure.

Before developing the technical framework, it is useful to situate this work within existing traditions. The problem of organisation has been approached from multiple directions, each exposing boundaries that remain unresolved when computation and physical process are treated as distinct.

\section{Theoretical context}

The framework developed here draws on several foundational lines of work that converge on the problem of organisation. Each addresses a different aspect of the same underlying question: how organised systems persist, transform, and encounter intrinsic limits. What distinguishes the present account is not the introduction of a new lineage, but the resolution of tensions exposed across these traditions.

C. H. Waddington reframed biology in terms of developmental organisation rather than static entities. Through concepts such as canalisation and the epigenetic landscape, he characterised biological systems as constrained dynamical processes whose robustness arises from organisation itself. Development, in this view, is shaped by the structure of permissible trajectories rather than by the execution of encoded instructions. While this work clarified the primacy of organisation, it did not treat organisation as a computational process, nor did it provide criteria for identifying computational limits in material systems.

Alan Turing established that formal systems capable of sufficient expressive power necessarily encounter undecidable behaviours. These limits arise from structure rather than from implementation details. In later work on morphogenesis, Turing demonstrated that ordered patterns can emerge from matter-matter interaction without representation or control. However, a separation remained between formal computation, where limits were rigorously defined, and physical organisation, which was analysed dynamically but not identified as computation itself.

John von Neumann addressed the problems of reliability, reproduction, and increasing complexity in systems composed of unreliable components. His work on self-reproducing automata recognised organisation as central and showed that reliability must emerge from structure rather than from component perfection. Nevertheless, computation continued to be framed in terms of logical architectures realised by physical systems, preserving a distinction between abstract description and material enactment.

Robert Rosen articulated a sustained critique of mechanistic and representational accounts of life. He argued that living systems are defined by organisation and closure rather than by material components or algorithmic description, and that models of such systems are necessarily incomplete. While Rosen identified the limits of formal representation, his account did not provide a physical or experimental framework for identifying computation with organisation.

Each of these contributions exposes a boundary. Waddington identifies the limits of instruction-based explanations of development; Turing identifies intrinsic limits of formal systems; von Neumann identifies the limits of component-level reliability; Rosen identifies the limits of mechanistic modelling. What remains unresolved across these accounts is the relation between organisation and computation as a physical process. The following sections develop a framework that addresses this gap.

\section{Organisation and constraint}

An organised system is defined here as one that maintains a bounded set of state variables whose future evolution depends on their current configuration. Organisation is not identified with a static arrangement at a given moment, but with persistence across time in the presence of transformation, exchange, and disturbance. A system is organised if it remains identifiable as the same organisation while its material constituents and internal states change.

This definition distinguishes organisation from aggregation. An aggregate may change continuously without preserving relational coherence, whereas an organised system preserves specific relations among its components. Organisation therefore concerns the maintenance of relations rather than the persistence of parts. The relevant criterion is not whether individual components endure, but whether the relations that constitute the organisation are re-established as the system evolves.

Structure is defined as the set of constraints that delimit which state transitions are admissible for the system. Constraints may arise from physical boundaries, chemical affinities, mechanical couplings, geometric configuration, or topological connectivity. They do not prescribe specific trajectories or outcomes. Instead, they define a space of possible trajectories within which the system may evolve.

Constraints are relational rather than local. They act by restricting combinations of states, not by directing individual components through explicit rules. Structure is therefore not reducible to a list of parts, parameters, or instructions. It is a property of the system as a whole, expressed through dependencies among variables and maintained through ongoing interaction.

Function is the realised dynamics within this constrained space. Function is not imposed externally and does not exist independently of structure. For any organised system, the set of functions it can realise is fully determined by its constraints, while those constraints are revealed only through the functions that are enacted. Structure and function are thus mutually defining rather than hierarchically ordered.

This mutual definition rules out instruction-based interpretations. Constraints delimit possibility without prescribing outcomes. Behaviour arises from interaction within constraint, not from the execution of predefined sequences. What appears as control or regulation corresponds to the stabilisation of relations, not the application of rules.

Constraints in organised systems are not static. They must be continuously enacted and maintained through energetic processes. Without ongoing energy exchange, constraints degrade and organisation collapses. Persistence therefore depends on the system's capacity to counteract dissipation by restoring the relations that define its structure. Organisation is thus inseparable from energy flow and time.

From this perspective, organisation is inherently dynamical. It cannot be identified solely from instantaneous measurements or static descriptions. Organisation must be characterised through the system's behaviour across time, particularly its response to perturbation and its capacity to restore relational coherence.

In what follows, organisation is treated as the necessary condition for computation, while computation is defined as the temporal enactment of such organisation. This distinction clarifies the conceptual roles of organisation and computation without separating them ontologically.

\section{Information as relational invariance}

Within the present framework, information is not treated as symbolic content, encoded message, or transferable signal. Information is defined structurally as a relational invariant: a difference that persists by influencing which future state transitions remain admissible for an organised system. Counterfactual structure is physically grounded as the space of admissible trajectories defined by constraint and energetic viability, and is accessed through structured perturbation and the recovery or loss of relational invariants.

This definition shifts the focus from representation to consequence. A difference counts as information only insofar as it alters what the system can do next. Differences that do not affect future admissibility are absorbed as noise, regardless of their magnitude or origin. Informational significance is therefore not intrinsic to a perturbation itself, but to the relation between that perturbation and the organisation it encounters.

Because admissible state transitions are defined by constraint, information is inseparable from structure. The same perturbation may be informationally significant for one organisation and irrelevant for another. Information is not carried through a system as an object or message; it is enacted locally through changes in relational possibilities. What matters is not what is received, but what becomes possible or impossible as a result.

Relational invariance should not be confused with static stability. Invariants may be preserved through recovery, compensation, or reconfiguration rather than exact repetition. Organised systems do not return to identical microstates after perturbation; they return to equivalent relational states. Information therefore concerns equivalence classes of organisation, not fixed configurations.

This view contrasts with transmission-based accounts of information in which information can be quantified independently of the system that receives it. In organised systems, informational content cannot be defined without reference to the constraints that shape future dynamics. Information is context-dependent in a precise structural sense: it is defined by its effect on admissibility.

Memory, within this framework, is the persistence of relational invariants across time. A system remembers insofar as previous states continue to constrain future behaviour. This persistence may be embodied in material configuration, spatial organisation, chemical gradients, mechanical stress patterns, or other structural features. Memory does not require the storage of symbols or records; it consists in the continued influence of past organisation on present dynamics.

Processing is the transformation of relational invariants under constraint. As the system evolves, some invariants are preserved, others are modified, and new invariants may be established. These transformations are continuous and distributed, not discrete operations applied to stored representations.

Execution is the forward propagation of these transformations into future states. Memory, processing, and execution are not separable stages but concurrent aspects of the same material process. Organised systems do not alternate between storing information, processing it, and acting upon it; these functions are realised simultaneously through ongoing interaction.

Information exists only insofar as organisation persists. When constraints degrade or collapse, relational invariants are lost and information ceases to exist for that system. Information loss is therefore not primarily a matter of signal degradation, but of organisational failure.

Methodologically, this definition implies that information cannot be identified solely through inputs or outputs. It must be inferred from the system's behaviour under perturbation: which relations are preserved, which are restored, and which fail. Informational content is revealed not by what passes through a system, but by how the system's future possibilities are reshaped.

By defining information as relational invariance, the framework aligns information with organisation rather than representation. Information is not an additional layer imposed on material dynamics; it is a property of organised matter insofar as it constrains its own future.

\section{Computation as enacted organisation}

Within the present framework, computation is not defined as the manipulation of symbols or the execution of externally specified instructions. A system computes insofar as it maintains and realises structured relations over time under constraint.

This definition follows from the preceding accounts of organisation and information. If structure defines admissible state transitions, and information consists in relational invariants that shape future admissibility, then computation is the process by which these invariants are preserved, transformed, and propagated through time. The warrant for computational vocabulary, rather than merely dynamical vocabulary, lies in this specific feature: computation occurs when a system's dynamics are constrained such that relational invariants persist or are restored. Mere change does not suffice. The system must exhibit structured persistence that shapes its own future admissible states. This is what distinguishes computation from dynamics more generally.

Computation is not added to organisation; it is the temporal expression of organisation.

This dependency can be expressed schematically. Let $R_t$ denote the system state at time $t$, $G$ the constraints defining admissible transitions, and $E$ the energetic flux sustaining those constraints. The system evolves according to
\[
R_{t+\Delta t} = \Phi(R_t; G, E).
\]
Computation occurs when relational invariants $I(R)$ are preserved or restored across this evolution,
\[
I(R_{t+\Delta t}) \approx I(R_t).
\]

An algorithm, in this sense, is not a discrete procedure encoded in a separate medium. It is an internally embedded regularity: a repeatable pattern of state transformation made possible by the system's constraints. What is traditionally described as an algorithmic step corresponds here to a transition that reliably occurs when the system occupies a particular region of its admissible state space. Algorithms are therefore enacted rather than executed.

This definition broadens the scope of what counts as algorithmic. Any organised system whose constraints enable repeatable state transformations possesses algorithms in this sense. This broadening is not a defect but a consequence of rejecting the separation between computation and physical process. The classical restriction of 'algorithm' to symbolic procedures reflected an assumption that computation is ontologically distinct from dynamics. Once that assumption is abandoned, there is no principled basis for denying algorithmic structure to non-symbolic systems. The selectivity of the framework lies not in which systems have algorithms, but in which systems compute: only those that maintain relational invariants under perturbation. A system may possess embedded regularities yet fail to compute if its organisation does not persist.

Under this definition, memory, processing, and execution are analytically distinguishable but physically inseparable aspects of the same material process.

This view contrasts with architectures in which computation is decomposed into discrete modules, such as memory units, processors, and control logic. In organised systems, such decomposition is not available. The same material structures that constrain behaviour also embody memory and enact change. There is no central locus of computation; computation is distributed across the relations that constitute the organisation.

Because computation is identified with organisation, it is inseparable from material conditions. Altering constraints alters what can be computed; altering energetic support alters whether computation can continue. Computation is therefore always situated, bounded, and contingent on the viability of the organisation that enacts it.

This perspective reframes classical notions of correctness and error. Correctness is not conformity to an external specification, but successful maintenance of organisation. Error is not a wrong output, but a failure to preserve relational invariants. Computation ends not when a program halts, but when organisation collapses.

By identifying computation with enacted organisation, the framework removes the distinction between computational process and physical process. Computation is not something that matter performs in addition to interacting; it is the manner in which organised matter persists through interaction.

\section{Hierarchy and limits}

When organised systems are composed of interacting subsystems, hierarchy arises as a structural consequence of organisation rather than as an imposed design. Subsystems operate over different spatial, temporal, and energetic scales, and their interactions lead to partial autonomy. Hierarchy, in this sense, does not consist in layered control or top-down instruction, but in the stabilisation of relations across scales.

At each scale, organisation is maintained locally through constraints that delimit admissible state transitions. These constraints are shaped by interactions with neighbouring subsystems and by slower, higher-level structures that change on longer timescales. As a result, higher-level organisation emerges from the persistence of lower-level relations while simultaneously constraining them. Hierarchy is therefore reciprocal: higher-level structures depend on lower-level dynamics, and lower-level dynamics are conditioned by higher-level constraints.

This reciprocal organisation entails that no component has access to the complete system state. Each subsystem operates on partial information defined by its position within the hierarchy and the timescales over which it can respond. Local processes cannot survey, represent, or control the organisation as a whole. Coordination arises through coupling and constraint rather than through global description.

Intrinsic limits follow directly from this structural arrangement. Because admissible state transitions are locally constrained and globally coupled, there exist behaviours that cannot be predicted, resolved, or controlled from within the system. These limits are not consequences of finite resources or insufficient precision; they arise from the organisation of the system itself.

This generalises classical results in computation theory, originating with Alan Turing, which show that sufficiently expressive systems contain undecidable behaviours. The claim here is not a formal reduction to symbolic undecidability, but a structural generalisation: whenever organisation combines self-reference with partial observability across scales, intrinsic limits on prediction and control necessarily arise. Hierarchy introduces limits not by complexity alone, but by coupling these two structural features.

In organised matter, such limits manifest as boundaries of viability. Beyond certain thresholds of perturbation, constraint alteration, or energetic depletion, relational invariants can no longer be restored. Organisation fails, and computation ceases. These boundaries are structural rather than algorithmic.
They cannot be removed by increasing speed, redundancy, or precision without altering the organisation itself.
Although organised dynamics are continuous and graded, their outcomes are discrete.  Organisation either persists or collapses; computation continues or ceases. This discreteness arises at viability boundaries, where continuous dynamics fail to sustain relational invariants, and the resulting organisational state conditions all subsequent evolution.

Hierarchy also constrains intervention. Because higher-level organisation is stabilised by interactions among lower-level processes, targeted manipulation of individual components may not yield predictable outcomes at the organisational level. Conversely, modifying constraints at higher scales can reshape the space of admissible lower-level behaviour without specifying particular trajectories. Control therefore operates indirectly, through constraint modulation rather than instruction.

From this perspective, computational limits are inseparable from organisational limits. What can be computed by an organised system is determined by the relations it can sustain and the scales over which those relations persist. Adding components or increasing activity does not necessarily expand computational capacity; it may destabilise existing constraints and reduce persistence.

Hierarchy thus functions simultaneously as an enabling condition for complex computation and as a source of intrinsic limitation. Organised systems compute by maintaining coherence across scales, and it is this same coherence that prevents complete self-description or total control. These limits are not external restrictions imposed on computation; they are consequences of organisation itself.

\section{What this framework excludes}

Because this framework defines computation structurally rather than metaphorically, it necessarily excludes several common interpretations of computation. These exclusions follow directly from the identification of computation with organised matter interacting under constraint.

This account does not endorse pancomputationalism. Physical systems may undergo change, exchange energy, or exhibit complex dynamics without computing in the sense defined here. Computation requires organised persistence: the maintenance or restoration of relational invariants under perturbation. Systems that dissipate without re-establishing organisational relations do not compute, regardless of their complexity or scale.

The framework excludes representational accounts of computation. There are no internal symbols that stand in for external states, no encoded data that are interpreted by a separate mechanism, and no distinction between data and process within the system itself. Meaning does not arise from correspondence between internal representations and external referents, but from the system's capacity to reshape its own future behaviour through maintained relations.

This account rejects instruction-based explanations. Organised systems do not execute programs in the sense of following predefined sequences of operations. Constraints delimit admissible transitions, but they do not prescribe outcomes. Behaviour arises from interaction within constraint, not from the application of rules.

The framework does not equate computation with dynamical change. Not all dynamics constitute computation. A system that changes without preserving or restoring relational structure performs dynamics but does not compute.

Finally, the framework does not claim that all biological phenomena reduce to computation, nor that computation replaces other forms of explanation. Instead, it provides criteria for when computation is present, how it degrades, and where its limits lie. These criteria impose boundaries on applicability rather than expanding the term to encompass all phenomena.

By making these exclusions explicit, the framework avoids conflating computation with description or metaphor. Computation is identified only where organisation persists through constrained interaction, and it ceases when that organisation can no longer be maintained.

\section{Experimental consequences}

Identifying computation with organised matter interacting under constraint entails that computation is not inferred from abstract input–output relations or task performance, but from the capacity of a system to maintain organisation through perturbation. Computation becomes experimentally accessible as a property of persistence, recovery, and failure rather than as correct execution of predefined operations. 

The primary experimental criterion is the persistence of relational invariants. An organised system computes insofar as relations that constrain future behaviour are preserved or restored following disturbance. These relations need not correspond to fixed configurations or repeatable trajectories. Instead, they define equivalence classes of organisation: distinct microstates that nevertheless support the same admissible transformations. Experimentally, this implies that repeated perturbations may produce variable immediate responses while converging on the same relational state.

A second criterion is recovery following structured perturbation. Perturbations are informative only when they challenge organisational constraints. By applying controlled disturbances—mechanical, chemical, geometric, or energetic—and observing whether relational invariants are re-established, one can determine whether computation continues or fails. Recovery is not defined by return to an initial state, but by restoration of admissible relations. The reliability and timescale of recovery provide quantitative measures of organisational viability.

A third criterion concerns phase boundaries in constraint–energy space. Organised systems depend on energetic flux to enact and restore constraints. For a given organisational structure, there exist ranges of energetic support within which relational invariants persist and ranges beyond which they cannot be maintained. By systematically varying constraints and energetic input, it is possible to map regions where computation is sustained and identify boundaries where organisation collapses.

Failure plays a central role in this framework. Loss of computation is marked by irreversible loss of relational invariants, not by incorrect outputs or halted execution. Experimental failure—where invariants cannot be restored following perturbation—is not an anomaly to be excluded, but a primary indicator of computational limits. The framework does not guarantee sharp classificatory boundaries in all cases: systems may exhibit transient or partial persistence without sustaining organisational invariants across scales, and such borderline cases are expected rather than pathological.

\subsection{Illustration: Chemotactic gradient sensing}

Consider the chemotactic signalling network in \textit{Dictyostelium discoideum}. This system maintains a relational structure linking receptor activation states, intracellular phosphorylation cascades, and motility output such that cells can climb chemical gradients. The organisation enables directional migration, a behaviour that depends on the persistence of relational invariants coupling receptor occupancy to motor bias.

Within this framework, the chemotactic network computes insofar as its relational structure persists through perturbation. We have recently shown that chemotactic sensitivity is not determined by the percentage difference in ligand concentration across the cell ($\Delta C\%$), but by the differential proportion of activated receptors ($\Delta \Omega$) (Dowdell et al., 2025). Cells can detect receptor-activation differences as small as $0.026\%$ across their length, corresponding to differentials of only two to three receptors at the cell periphery. Crucially, the minimum differential required to induce directional bias increases with overall receptor activation: in saturated environments, cells become less sensitive to directional information.

This system illustrates the experimental criteria of the framework directly.

\emph{Persistence of relational invariants.} The coupling between receptor activation and directional output constitutes a relational invariant. Across four orders of magnitude in ligand concentration, this coupling persists provided that $\Delta \Omega$ exceeds a threshold. The invariant is not the ligand concentration itself, nor any particular receptor configuration, but the relational structure linking differential activation to behavioural output.

\emph{Phase boundaries in constraint space.} The active-receptor (AR) model predicts, and experiments confirm, distinct regions in which chemotaxis is sustained, weakened, or fails entirely. In steep gradients (for example, source concentrations of $100$--$300\,\mu\text{M}$), cells exhibit strong initial directional bias that collapses as receptor saturation eliminates the differential signal. In shallow gradients (for example, $1\,\mu\text{M}$), sustained bias persists across most of the gradient. The position $x_c$ at which chemotaxis ceases marks a phase boundary: beyond this point, the organisational relation linking receptor state to motor output can no longer be maintained. These boundaries can be mapped by systematically varying gradient steepness and identifying where directional behaviour fails.

\emph{Failure as an indicator of computational limits.} The cessation of chemotaxis at $x_c$ is not a performance error but an organisational failure. The relational invariant---differential receptor activation sufficient to bias movement---has been lost. The framework predicts that such limits are structural rather than algorithmic: they arise from the organisation of the receptor--signalling--motor system itself and cannot be circumvented by increasing speed or precision without altering the underlying constraints.

\subsection{Reframing experimental design}

This example reframes experimental design. Rather than evaluating performance on predefined tasks, experiments should probe organisational resilience. Relevant observables include the maintenance of spatial, mechanical, chemical, or dynamical relations; the system's capacity to re-establish these relations after perturbation; and the dependence of this capacity on constraint and energy. Computation is inferred from sustained organisation, not from functional output alone. 

For dynamic systems like chemotactic networks, this means identifying the relational invariants that couple input to output and mapping the perturbation regimes under which these invariants persist or fail.

This framework applies across scales and substrates. It does not require biological specificity, nor does it depend on digital architecture. Any system in which constraints can be identified, perturbations applied, and recovery assessed is amenable to this analysis. Conversely, systems that fail to restore organisation under perturbation do not compute, regardless of their apparent complexity or functionality.

By grounding computation in experimentally observable persistence and failure, this framework transforms computation from a descriptive attribution into a testable property of organised matter.

\section{Limitations}

Several limitations constrain the scope and application of this framework.

The framework does not specify how to individuate organisations. Boundaries between one organised system and another, or between an organisation and its environment, are not given by the framework itself. In practice, such boundaries must be determined operationally through perturbation and response, which introduces dependence on experimental choice.

The criterion of relational invariance admits degrees. Systems may exhibit partial or transient persistence without clearly falling on one side of a computational/non-computational boundary. The framework acknowledges such borderline cases but does not resolve them into sharp categories.

The relationship between this framework and existing computational formalisms remains underdeveloped. While the account claims to generalise classical results from computability theory, no formal reduction is provided. Whether the structural limits described here can be made rigorous in terms comparable to Turing's results is an open question.

The framework does not engage systematically with related contemporary approaches, including integrated information theory, constructor theory, or enactivist accounts of cognition. Such engagement would clarify points of convergence and divergence but lies beyond the present scope.

Finally, the experimental criteria proposed here have not yet been applied systematically. The framework provides a conceptual reorientation, but its empirical utility remains to be demonstrated through sustained experimental programmes.

\section{Conclusion}

This paper has developed a structural definition of computation in which computation is identified with organised matter interacting under constraint. By removing the separation between logical description and physical realisation, the framework treats structure, function, information, memory, and algorithm as mutually defining aspects of a single process.

Within this account, information is not transmitted content or encoded representation, but relational invariance defined by its influence on future admissible state transitions. Memory consists in the persistence of such invariants, processing in their transformation under constraint, and execution in their continuation into future states. Computation is therefore not the manipulation of symbols, but the maintenance and enactment of admissible relations in organised matter.

This identification situates the limits of computation within organisation itself. Hierarchy, partial observability, and self-reference generate intrinsic boundaries on what organised systems can sustain or resolve. These limits generalise classical results from computability theory to material systems without requiring symbolic formalism as a prerequisite.

By making computation contingent on persistence, the framework avoids both representational and pancomputational accounts. The computational framing is retained not as metaphor, but because it exposes intrinsic limits, hierarchy, and failure modes that are not captured by descriptions of organisation alone.

Crucially, this view renders computation experimentally accessible. Persistence, recovery, and structural failure provide criteria for identifying when computation occurs, how it degrades, and where its limits lie. Computation can be sustained, degraded, or lost, and its boundaries can be mapped by varying constraint and energetic support.

The framework opens several directions for further work. Empirically, systematic application of the perturbation-recovery methodology across biological and synthetic systems would test its utility. Theoretically, formalising the relationship between organisational limits and classical undecidability results would strengthen the account. Comparatively, engagement with related frameworks---integrated information theory, constructor theory, enactivism---would clarify the distinctive contributions and boundaries of this approach.

The framework does not propose a new metaphor for life, nor does it replace existing explanatory approaches. Instead, it offers a structural account in which computation is neither imposed on matter nor abstracted away from it. Computation is identified with organisation itself, insofar as organised matter persists through interaction.

\section*{Acknowledgements}
Supported by the ARIA “Trust Everything, Everywhere” Opportunity Space; \textit{Living Proof}; and the “Nature Computes Better” Opportunity Space, \textit{(Bio)active Matter Based Computation}.

\section*{References}

\begin{itemize}
  \item Waddington, C.~H. (1957) \textit{The Strategy of the Genes: A Discussion of Some Aspects of Theoretical Biology}. London: George Allen \& Unwin.

  \item Turing, A.~M. (1936) `On computable numbers, with an application to the Entscheidungsproblem', \textit{Proceedings of the London Mathematical Society}, Series~2, 42(1), pp.~230--265.\\
  \texttt{https://doi.org/10.1112/plms/s2-42.1.230}

  \item Turing, A.~M. (1952) `The chemical basis of morphogenesis', \textit{Philosophical Transactions of the Royal Society of London. Series~B, Biological Sciences}, 237(641), pp.~37--72.\\
  \texttt{https://doi.org/10.1098/rstb.1952.0012}

  \item von Neumann, J. (1966) \textit{Theory of Self-Reproducing Automata}. Edited by A.~W. Burks. Urbana, IL: University of Illinois Press.

  \item Rosen, R. (1991) \textit{Life Itself: A Comprehensive Inquiry into the Nature, Origin, and Fabrication of Life}. New York: Columbia University Press.
  
  \item Dowdell, A., Jordan, D., and Witte, K. (2025). \textit{Differential receptor activation defines the fundamental limit to chemotactic sensitivity}. \textit{bioRxiv}. https://doi.org/10.64898/2025.12.17.694457
\end{itemize}

\end{document}